\documentstyle[12pt]{article}
\vspace{5mm}
\title{NEW EQUATIONS FOR VACUUM CORRELATORS IN GAUGE THEORIES}
\vspace{2mm}
\author{D.V.Antonov \\
Institute of Theoretical and
Experimental Physics,\\
B.Cheremushkinskaya, 25, 117218, Moscow, Russia}
\date{}
\newcommand{\be}{\begin{equation}}
\newcommand{\ee}{\end{equation}}

\begin{document}
\maketitle
\large

\vspace{10mm}
\centerline{\bf Abstract}

Stochastic quantization [1] is applied to derivation of equations,
connecting multilocal gauge-invariant correlators in different field
theories. They include  Abelian Higgs Model, QCD with spinless quarks at $T
= 0$ and $T > 0$ and QED, where spin effects are taken into account exactly.

\vspace{7mm}
\section{Introduction}

\vspace{3mm}
\hspace{5mm} One of the main problems of the modern quantum field theory is
determination of nonperturbative effects in the models with nontrivial
vacuum structure, like QCD. To investigate  this problem the Method of
Vacuum Correlators (MVC) [2-8, 17]  (for a review see [5]) was suggested. In
the framework of this method one, using Feynman-Schwinger path integral
representation, extracts from quark-antiquark Green function all the
dependence from\\
vacuum gluonic fields in the form of an averaged Wilson
loop. After that, through nonabelian Stokes theorem [3] and cumulant
expansion [3,9] the ave-\\
raged Wilson loop is expressed via an infinite set
of irreducible gauge-invariant vacuum correlators (cumulants). Lattice data
suggest, that the ensemble of gluonic fluctuations is predominantly Gaussian
[5]. That is why, in order to calculate observable quantities (like
confining potential [4,5], mass spectra of mesons, baryons and glueballs
[5-7], Regge trajectories [7,8], temperature of deconfinement, meson and
glueball screening masses at finite temperature [17], etc.) one may with a
good accuracy neglect all the cumulants higher than quadratic and to
parametrize the latter by two independent Kronecker structures, where the
space-time dependence of the corresponding coefficient functions is
determined from the lattice data.

However, the nonperturbative stochastic Euclidean QCD vacuum is described by
the full set of cumulants, and, therefore, for the completeness of the
theory, one needs the equations (derived from the Lagrangian), from which
the cumulants may be obtained. Two alternative
methods of derivation of such equations, based on stochastic quantization
[1], were suggested in [10].  First of them, exploiting Feynman-Schwinger
path integral representation, where Langevin time is considered as a proper
time, was applied to the $\varphi^3$ theory. However, it is inconvenient for
gluodynamics, since it breaks down gauge invariance. That is why, in the
latter case an alternative approach, leading to integral-differential
Volterra type-II equations, was used. In both cases it was shown, that the
perturbative expansion of the obtained equations in the lowest orders
reproduced correctly known results of standard Feynman diagrammatic
technique. An important point is that in the physical limit, when Langevin
time $t$ tends to infinity, the solutions of these equations yield exact
correlators of Heisenberg field operators, averaged over the physical vacuum,
whose structure may be complicated and unknown, e.g. in QCD.

In this paper we demonstrate, how these two approaches work simultaneously
in gauge theories with matter fields, presenting the methods of derivation
of equations for gauge-invariant correlators in different models. The sketch
of the paper is the following. In section 2 we start with the Abelian Higgs
Model (AHM), where classical solutions (Abrikosov-Nielsen-Olesen (ANO)
strings) are present. These objects may be naturally included into
obtained equations as initial conditions in the Cauchy problem for the
Langevin equation, and, thus, we obtain a new method of quantization of
classical solutions. This method allows one to derive correlators,
containing arbitrary number of quantum fluctuations around classical
solutions in any theory. We hope, that it will be especially useful in
the case of gauge theories, since the obtained equations are
explicitly gauge-invariant. Computation of quantum corrections to an
instanton in the framework of this method will be a topic of a
separate publication. Another point is that the suggested approach is the
first attempt to gauge-invariant quantization of theories with spontaneous
symmetry breaking and may be applied to quantization of more complicated
theories, like Standard Model. In section 3 we exploit methods of stochastic
quantization of fermionic theories to derive equations for gauge-invariant
correlators in QCD with spinless quarks. Then we demonstrate, how the
obtained equations may be generalized to the finite temperature case.
Finally, in section 4 we use the path integral approach to calculation of the
propagator of a spinning particle in the Abelian background gauge field
[11,12] to obtain equations in QED, where all spin effects are taken into
account exactly. In Conclusion we summarize the main results of the paper
and outline possible future developments. In the Appendix we obtain a Green
function of the Langevin equation for $F_{\mu \nu}(x, t)$ in the AHM.

\vspace {3mm}
\section{Abelian Higgs Model}

\hspace{5mm} In this section we consider AHM,
where  exists a vacuum of classical
solutions (ANO strings) [13]. The Euclidean action of the AHM has the form
\be
S = \int~ dx \left [(D_{\mu} \varphi)^* (D_{\mu}\varphi) + \frac{1}{4} F_{\mu
\nu}F_{\mu \nu} - \lambda(\varphi^* \varphi)^2 + m^2 \varphi^* \varphi
\right ],
\ee
where $D_{\mu} \varphi = (\partial_{\mu} - ie A_{\mu}) \varphi$ .
Let us look for a solution of equations of motion in the cylindrical
coordinates in the form $\vec{A}(\vec{r}) = A_{\theta}(r), ~ A_4 = 0,
\varphi = \mid \varphi \mid e^{in \theta}$, where $\mid \varphi \mid$ is a
function of $r$ only. Then it is easy to check, that such a solution (ANO
string along z-axis), corresponding to an $n$-quantum Abrikosov vortex of
Ginzburg-Landau model, has the form
\be
A_{\theta}(r) = \left\{
\begin{array}{ll} \frac{n}{er} - \frac{n}{e \delta} K_1(\rho),& r \geq
\delta,\\
\frac{n}{2 e \delta} (-\frac{1}{2} + \frac{1}{2} \rho - \rho ln
   \rho), & \xi \ll r \ll \delta,
   \end{array}
   \right.
\ee
where $\delta \equiv \frac {\sqrt \lambda}{me}$ is a Londons$'$ depth of
penetration of the field $A_{\mu},~ \xi \equiv \frac{1}{m}$ is a correlation
radius of fluctuations of the Higgs field, $\rho \equiv \frac{r}{\delta},
K_1$ is the Macdonald function.

The Langevin equations, following from the action (1), have the form

\be
\dot{\varphi} = D^2 \varphi - m^2 \varphi + 2 \lambda \varphi^* \varphi^2 +
\eta,
\ee

\be
\dot{\varphi}^* = (D^2 \varphi)^* - m^2 \varphi^* + 2 \lambda \varphi
\varphi^{* 2} + \eta^*,
\ee

\be
\dot{A}_{\mu} = \partial_{\nu} F_{\nu \mu} + j_{\mu} +
\eta_{\mu},
\ee
where $j_{\mu} = -ie (\varphi^* \partial_{\mu} \varphi - (\partial_{\mu}
\varphi^*) \varphi - 2 ieA_{\mu} \varphi^* \varphi)$ is a gauge-invariant
current,
$$ <\eta(x, t) \eta^*(x^{\prime}, t^{\prime}) > =2 \delta(x -
x^{\prime}) \delta(t - t^{\prime}), ~~~
$$
\be
<\eta_{\mu} (x, t)
\eta_{\nu}(x^{\prime}, t^{\prime}) > = 2\delta_{\mu \nu} \delta(x -
x^{\prime})\delta(t - t^{\prime}),
\ee
and $< ...>$ means the averaging
over both Gaussian stochastic ensembles of $(\eta, \eta^*)$ and
$\eta_{\mu}$.

Passing from equation (5) to the Langevin equation for $F_{\mu \nu}(x, t)$
 and \\
 solving it via Fourier transformation with the initial condition
$A_{\mu}(x, 0)=\\
=A_{\mu}(x)$, where $A_r(x)=A_z(x)=A_4(x)=0$ and
 $A_{\theta}(x)$ is defined through the formula (2), one obtains:
$$
F_{\mu \nu}(x, t) = \int\limits_{0}^{t}~ dt^{\prime}~\int~dy~U_{\mu
 \nu,\beta}(y-x,t - t^{\prime}) (j_{\beta}(y,t^{\prime}) +
 \eta_{\beta}(y,t^{\prime})) + $$
\be
 +\int~ dy~ U_{\mu \nu,\beta}(y-x,t)
 A_{\beta}(y),
\ee
where
$$
U_{\mu \nu, \beta}(y,t) = \frac{1}{48 \pi^2} \left \{(\delta_{\mu \alpha}~
\delta_{\nu \beta} - \delta_{\nu \alpha}~ \delta_{\mu \beta})
\Biggl [\frac{y_{\alpha}}{t^3} e^{-\frac{y^2}{4t}} - 16
\pi^2 \Biggl (\frac{\partial}{\partial y_{\alpha}}~\delta(y) \Biggr)
\Biggr]+ \right.$$

$$ + \frac{2}{y^2}(\delta_{\nu \beta}y_{\mu} -
\delta_{\mu \beta}y_{\nu}) \Biggl( \frac{1}{t^2}e^{-\frac{y^2}{4t}} -
16\pi^2 \delta(y) \Biggr ) + \frac{1}{y^2} \Biggl ( \delta_{\mu \beta}
y_{\nu} y_{\alpha} +$$

$$ + \delta_{\nu \alpha} y_{\mu} y_{\beta}
-\delta_{\nu \beta} y_{\mu} y_{\alpha} -\delta_{\mu \alpha} y_{\nu}
y_{\beta} \Biggr )\cdot $$

\be
\left. \cdot \Biggl [\frac{y_{\alpha}}{2t^3}
 e^{-\frac{y^2}{4t}} + 16 \pi^2 \Biggl (\frac{\partial}{\partial
 y_{\alpha}}~\delta(y)\Biggr )\Biggr ] \right \}.
\ee
The details of derivation of the Green function (8) are presented in the
Appendix .

Using Langevin time as a proper time and solving equation (3) via
Feynman-Schwinger path integral representation with the initial condition
$\varphi(x,0) = \varphi_0(x), \mid\varphi_0 \mid^2 = \frac{m^2}{2 \lambda}$,
we have:

\be
\varphi(x,t) = \int \limits_{0}^{t}~ dt^{\prime}~ \int~
dy(Dz)_{xy} K_z(t, t^{\prime}) {\cal{F}}_z(t, t^{\prime}) \Phi(x,y) \eta(y,
t^{\prime}),
\ee
where
$$
(Dz)_{xy} \equiv \lim_{N \rightarrow \infty} \prod
\limits _{n=1}^{N}~ \frac{d^4 z(n)}{(4 \pi \varepsilon)^2}, ~~~, N
\varepsilon = t-t^{\prime}, ~~ z(\xi=t^{\prime})=y, ~~ z(\xi=t) = x,
$$

$$
K_z(t, t^{\prime}) \equiv \theta(t-t^{\prime}) exp \left [-m^2(t-t^{\prime})
-\int \limits_{t^{\prime}}^{t}~ \frac{\dot{z}^2(\xi)}{4} d\xi, \right ],$$

$${\cal{F}}_z(t,t^{\prime}) \equiv exp
                            \left [2 \lambda
\int \limits_{t{^\prime}}^{t}~
d \xi \mid \varphi(z(\xi), \xi) \mid^2 \right],$$

$$
\Phi(x,y) \equiv exp \left [ ie~ \int \limits _{y}^{x}  dz_{\mu}
A_{\mu} (z(\xi), \xi)
 \right ]$$
and we omitted in (9) the term $\int dy (Dz)_{xy} K_z(t,0) {\cal{F}}_z (t,0)
\Phi(x,y) \varphi_0(y)$ , because we shall see, that in the correlators (11),
(12) (and all the other gauge-invariant quantities) this term produces
non-gauge-invariant part of the correlator, which in the physical limit, $t
\rightarrow +~ \infty$, tends to zero.

In the same way one may solve equation (4).

To obtain an infinite system of exact equations for gauge-invariant
correlators, containing the fields
                               $\varphi$ and $\varphi^*$, one should
 multiply the right hand sides of equation (9) and the equation for
 $\varphi^*$ by $\eta(\bar{x}, \bar{t}),~ \eta^*(\bar{x}, \bar{t}),~
 \eta_{\mu}(\bar{x}, \bar{t}),~ \varphi(\bar{x}, \bar{t}),~
 \varphi^*(\bar{x}, \bar{t}),~ j_{\mu}(\bar{x}, \bar{t}) ,~ F_{\mu
 \nu}(\bar{x},\bar{t})$ as many times as needed, introducing corresponding
 parallel transporters in order to preserve gauge invariance, and average
 both sides of the obtained equations.  The right hand sides of these
 equations may be most easily expressed through their left hand sides by
 introducing the generating functional
 $$
 Z = \prod\limits_{i=1}^{k} <
 {\cal{F}}_{z_i}(t_i,t^{\prime}_i) W_i(C_i) exp \Biggl [\int~ du dt (I_{\mu}
 (u,t) \eta_{\mu} (u,t)+$$

  \be
     +K_{\mu} j_{\mu} + L_{\mu \nu} F_{\mu \nu} + M \eta
+ N \eta^*)\Biggr ] >,
\ee
where $k$ is a total number of fields $\varphi$
and $\varphi^*$ in the correlator, $W_i(C_i)$ is an Abelian Wilson loop, and
the form of a contour $C_i$ is clarified below. After that one should apply
to this  generating functional cumulant expansion [3,9].

For example, let us consider the simplest case $k = 2$. This is the
so-called bilocal approximation [10], when one assumes, that the ensemble of
the fields $\varphi, \varphi^*, j_{\mu}, F_{\mu \nu}$ and stochastic noise
fields is Gaussian, so that all the cumulants, higher than quadratic, are
equal to zero. In this case we obtain
$$
<\varphi^*(\bar{x}, \bar{t}) \Phi(\bar{x}, x, \tau) \varphi(x, t) > =
$$
\be
= 2\int\limits_{0}^{min(t, \bar{t})}dt^{\prime} \int dy(Dz)_{xy}
(D\bar{z})_{\bar{x}y}K_z(t, t^{\prime}) K_{\bar{z}} (\bar{t},t^{\prime})
 < {\cal{F}}_z(t,t^{\prime}){\cal{F}}_{\bar{z}}(\bar{t}, t^{\prime})
W_{y\bar{x}xy} >,
\ee
$$
< \eta^*
(\bar{x}, t^{\prime}) \Phi(\bar{x}, x, \tau) \varphi(x,t) > = -<
\varphi^*(x,t) \Phi(x, \bar{x}, \tau) \eta(\bar{x}, t^{\prime}) > =
$$
\be
= 2
\int (Dz)_{x \bar{x}} K_z(t, t^{\prime}) < {\cal{F}}_z (t, t^{\prime})
W_{\bar{x} x \bar{x}}>,
\ee
where $\Phi(\bar{x},x, \tau)$ is a parallel
transporter between the points $x$ and $\bar{x}$, given at the moment $\tau$
of Langevin time (in particular, one may consider the case, when $x =
\bar{x}$ and $\Phi(\bar{x},x, \tau) = 1)$.  Here
\be
W_{y\bar{x}xy}
\equiv exp \left \{ie \Biggl [\int\limits_{\bar{x}}^y~d\bar{z}_{\mu} A_{\mu}
(\bar{z}(\bar{\xi}),\bar{\xi}) + \int\limits_{x}^{\bar{x}} du_{\mu}
A_{\mu}(u, \tau) + \int\limits_{y}^{x} dz_{\mu} A_{\mu}(z(\xi), \xi)
\Biggr ]\right \},
\ee
\be
W_{\bar{x}x\bar{x}} \equiv exp \left \{
ie~\Biggl [~\int\limits_{x}^{\bar{x}} du_{\mu} A_{\mu}(u, \tau) +
\int\limits_{\bar{x}}^{x} dz_{\mu}A_{\mu} (z(\xi), \xi) \Biggr ] \right \}.
\ee
Formulae (13) and (14) clarify the construction of the contours $C_1$ and
$C_2$ in bilocal
generating functional. Applying to it  cumulant
expansion, varying over $M$ or $N$ in the case of equations (12), putting
$I_{\mu} = K_{\mu} = L_{\mu \nu} = M = N = 0$ and rewriting $W_{y\bar{x}xy}$
and $W_{\bar{x}x\bar{x}}$ through Stokes theorem, one obtains in the bilocal
approximation:
$$
< {\cal{F}}_z(t,t^{\prime}){\cal{F}}_{\bar{z}}(\bar{t},
t^{\prime})W_{y\bar{x}xy} > = exp \Biggl [2 \lambda
\int\limits_{t^{\prime}}^{t} d\xi < \mid \varphi(z(\xi), \xi) \mid^2 >+$$

$$+ 2\lambda \int\limits_{t^{\prime}}^{\bar{t}} d\bar{\xi} < \mid
\varphi(\bar{z}(\bar{\xi}), \bar{\xi}) \mid^2 > + $$

\be
+ \frac{1}{2}
\int\limits_{S_1} d\sigma_{\mu_1 \nu_1}(u_1) \int\limits_{S_1}
d\sigma_{\mu_2 \nu_2}(u_2) < F_{\mu_1 \nu_1}(u_1, t_1)F_{\mu_2 \nu_2}(u_2,
t_2) > \Biggr ],
\ee

$$
<{\cal{F}}_z(t,t^{\prime})W_{\bar{x}x\bar{x}} > = exp
\Biggl [2\lambda \int\limits_{t^{\prime}}^{t} d\xi < \mid \varphi(z(\xi),
 \xi) \mid^2 > +   $$

 \be
 +\frac{1}{2} \int\limits_{S_2} d\sigma_{\mu_1 \nu_1}(u_1)
 \int\limits_{S_2} d\sigma_{\mu_2 \nu_2}(u_2) < F_{\mu_1\nu_1}(u_1,t_1)
 F_{\mu_2 \nu_2}(u_2, t_2) > \Biggr ],
 \ee
 where $S_1$ and $S_2$
are arbitrary surfaces, bounded by the contours $C_1$ and $C_2$
respectively, $t_1$ and $t_2$ are some moments of Langevin time.

The solution (7) of the Langevin equation for $F_{\mu \nu}(x,t)$ yields
three more equations for gauge-invariant correlators:
$$
<F_{\mu \nu}(x,t)F_{\lambda \rho}(\bar{x}, \bar{t}) > = \int\limits_{0}^{t}
dt^{\prime} \int\limits_{0}^{\bar{t}}d\bar{t}^{\prime} \int dy
d\bar{y}~U_{\mu \nu,\alpha}(y-x,t-t^{\prime}) U_{\lambda \rho,
\beta}(\bar{y}-\bar{x}, \bar{t}-\bar{t}^{\prime}) \cdot
$$

$$
\cdot \Biggl [ <
j_{\alpha}(y, t^{\prime})j_{\beta}(\bar{y}, \bar{t}^{\prime}) > +<
j_{\alpha}(y,t^{\prime})\eta_{\beta}(\bar{y}, \bar{t}^{\prime})> + <
j_{\beta}(\bar{y},\bar{t}^{\prime}) \eta_{\alpha}(y,t^{\prime}) > +
$$

$$
+
2\delta_{\alpha \beta}(y-\bar{y})\delta(t^{\prime}-\bar{t}^{\prime})
\Biggr ] +
$$
\be
+ \int~ dy d\bar{y} U_{\mu \nu, \alpha}(y-x,t) U_{\lambda
\rho, \beta}(\bar{y}-\bar{x},\bar{t}) A_{\alpha}(y) A_{\beta}(\bar{y}),
\ee
$$
<F_{\mu \nu}(x,t)j_{\lambda}(\bar{x}, \bar{t}) > =
\int\limits_{0}^{t} dt^{\prime}\int dy~U_{\mu \nu, \alpha}(y-x,t-t^{\prime})
\Biggl [ <j_{\alpha}(y,t^{\prime})j_{\lambda}(\bar{x},\bar{t})> +
$$
\be
+ <j_{\lambda}(\bar{x},\bar{t})\eta_{\alpha}(y, t^{\prime})> \Biggr ],
\ee
$$
<F_{\mu \nu}(x,t)\eta_{\lambda}(\bar{x}, \bar{t}) >
= \int\limits_{0}^{t} dt^{\prime}\int dy~U_{\mu \nu,
\alpha}(y-x,t-t^{\prime})
\Biggl [ <j_{\alpha}(y,t^{\prime})\eta_{\lambda}(\bar{x}, \bar{t})>+
$$

\be
+
2\delta_{\alpha \lambda} \delta(y-\bar{x}) \delta(t^{\prime}
-\bar{t}) \Biggr].
\ee

In order to obtain a complete system of equations, let us pass from the
Hamilton gauge to the Schwinger gauge, $A_{\mu}(x,t)(x-x_0)_{\mu} = 0$
(where $x_0$ is an arbitrary point), in which it is easy to express
explicitly $A_{\mu}(x,t)$ through $F_{\mu \nu}(x,t)$ and to obtain
gauge-invariant equations. One can see, that a gauge function,
corresponding to such gauge transformation, should depend
on $\theta$ and $z$ only. After that we introduce the second
generating functional
\be
\Phi_{\beta}(t) = exp \left \{ ie \oint\limits_C dx_{\mu} \left [
\int\limits_{x_0}^{x} dz_{\nu} \alpha(z,x) F_{\nu \mu}(z,t) + \beta
\int\limits_{0}^{t} dt^{\prime}(j_{\mu}(x,t^{\prime}) +
\eta_{\mu}(x,t^{\prime})) \right ] \right \},
\ee
where $\beta$ is a $c$-number, $\alpha(z,x) \equiv
\frac{(z-x_0)_{\nu}}{(x-x_0)_{\nu}}, C$ is some fixed closed contour.
Differentiating $\Phi_{\beta}(t)$ by $t$, applying to it cumulant expansion,
differentiating once by $\beta$ and putting then $\beta$ equal
to $(-1)$, we obtain, due to (5), the two last equations of bilocal
approximation:
$$
\frac{1}{2} \left \{\int\limits_{x_0}^{y} dz_{\lambda} \alpha(z,y)
\int\limits_{x_0}^{u} dx_{\rho} \alpha(x,u) \frac{\partial}{\partial t} <
F_{\lambda \nu}(z,t) F_{\rho \mu}(x,t) > - \right.$$

$$ - \int\limits_{x_0}^{y}dz_{\lambda}
\alpha(z,y) \int\limits_{0}^{t} dt^{\prime} \frac{\partial}{\partial t}
\Biggl [ < F_{\lambda \nu}(z,t)  j_{\mu}(u,t^{\prime}) > + < F_{\lambda
\nu}(z,t)\eta_{\mu}(u,t^{\prime}) > \Biggr ] -
$$

$$
- \int\limits_{x_0}^{u} dx_{\rho} \alpha(x,u) \int\limits_{0}^{t}
dt^{\prime} \frac{\partial}{\partial t} \Biggl [ < F_{\rho \mu}(x,t)
j_{\nu}(y,t^{\prime}) > + < F_{\rho \mu}(x,t) \eta_{\nu}(y,t^{\prime}) >
\Biggr ] -
$$

$$
- \int\limits_{x_0}^{y} dz_{\lambda} \alpha(z,y) \Biggl [ < F_{\lambda
\nu}(z,t) j_{\mu}(u,t) > + < F_{\lambda \nu}(z,t) \eta_{\mu}(u,t) > \Biggr ]
- $$

$$
- \int\limits_{x_0}^{u} dx_{\rho} \alpha(x,u) \Biggl [ < F_{\rho \mu}(x,t)
j_{\nu}(y,t) > + < F_{\rho \mu}(x,t) \eta_{\nu}(y,t) > \Biggr ] +
$$

$$
+ \int\limits_{0}^{t} dt^{\prime} \Biggl  [ < j_{\nu}(y,t)
j_{\mu}(u,t^{\prime}) > + < j_{\nu}(y,t^{\prime}) j_{\mu}(u,t) >+
< j_{\nu}(y,t) \eta_{\mu}(u,t^{\prime}) > +
$$

$$
\left.+ < j_ {\nu}(y,t^{\prime}) \eta_{\mu}(u,t) > + <j_{\mu}(u,t)
\eta_{\nu}(y,t^{\prime}) > + < j_{\mu}(u,t^{\prime}) \eta_{\nu}(y,t) >
 \Biggr ] \right \} + $$

 $$ + \delta_{\mu \nu} \delta(y-u) =
\frac{\partial}{\partial u_{\rho}} \left \{ \int\limits_{x_0}^{y}
dz_{\lambda} \alpha(z,y) < F_{\rho \mu}(u,t) F_{\lambda \nu}(z,t) > \right. -
$$

\be
- \int\limits_{0}^{t} dt^{\prime} \left.\Biggl [ < F_{\rho \mu}(u,t)
j_{\nu}(y,t^{\prime}) > + < F_{\rho \mu}(u,t)
\eta_{\nu}(y,t^{\prime}) > \Biggr ] \right \},
\ee
$$
\int\limits_{x_0}^{y} dz_{\lambda} \alpha(z,y) \frac{\partial}{\partial t}
\Biggl [ < F_{\lambda \nu}(z,t) j_{\mu}(u,t^{\prime}) > + <F_{\lambda
\nu}(z,t) \eta_{\mu}(u,t^{\prime}) > \Biggr ] - $$

$$- <j_{\nu}(y,t)j_{\mu}(u,t^{\prime}) >  - < j_{\nu}(y,t)
\eta_{\mu}(u,t^{\prime})> -$$

$$-< j_{\mu}(u,t^{\prime}) \eta_{\nu}(y,t) > + 2
\delta_{\mu \nu} \delta(y-u) \delta(t-t^{\prime}) -
$$

$$
- \frac{e^2}{2} \oint\limits_{C} dv_{\xi} \oint\limits_{C}dw_{\sigma}
\left \{ \int\limits_{x_0}^{v} d\bar{z}_{\lambda} \alpha(\bar{z}, v)~
\int\limits_{x_0}^{w} d\bar{x}_{\rho} \alpha(\bar{x}, w)
\frac{\partial}{\partial t} < F_{\lambda \xi}(\bar{z}, t) F_{\rho
\sigma}(\bar{x}, t) >- \right.  $$

$$
- \int\limits_{x_0}^{v} d \bar{z}_{\lambda} \alpha (\bar{z}, v)~
\int\limits_{0}^{t} d \bar{t}^{\prime} \frac{\partial}{\partial t} \Biggl [ <
F_{\lambda \xi}(\bar{z}, t)j_{\sigma}(w, \bar{t}^{\prime}) > + <
F_{\lambda \xi}(\bar{z}, t) \eta_{\sigma} (w, \bar{t}^{\prime}) > \Biggr ] -
$$

$$
- \int\limits_{x_0}^{w} d\bar{x}_{\rho} \alpha(\bar{x}, w)
\int\limits_{0}^{t} d\bar{t}^{\prime} \frac{\partial}{\partial t} \Biggl [ <
 F_{\rho \sigma}(\bar{x}, t) j_{\xi} (v, \bar{t}^{\prime}) > + < F_{\rho
\sigma}(\bar{x}, t)\eta _{\xi}(v, \bar{t}^{\prime}) > \Biggr ] -
$$

$$
- \int\limits_{x_0}^{v} d\bar{z}_{\lambda} \alpha(\bar{z},v) \Biggl [ <
F_{\lambda \xi}(\bar{z},t) j_{\sigma}(w,t) > + < F_{\lambda
\xi}(\bar{z},t) \eta_{\sigma}(w,t) > \Biggr ] -
$$

$$
-\int\limits_{x_0}^{w} d\bar{x}_{\rho} \alpha(\bar{x}, w) \Biggl [ <
F_{\rho \sigma}(\bar{x},t) j_{\xi}(v,t) > + < F_{\rho
\sigma}(\bar{x},t) \eta_{\xi}(v,t) > \Biggr ] +
$$

$$
+ \int\limits_{0}^{t} d\bar{t}^{\prime} \Biggl [
< j_{\xi}(v,\bar{t}^{\prime}) j_{\sigma}(w,t) > + <
j_{\xi}(v,t) j_{\sigma}(w,\bar{t}^{\prime}) > + <
j_{\xi}(v,\bar{t}^{\prime}) \eta_{\sigma}(w,t) > +
$$

$$
 + <j_{\xi}(v,t)\eta_{\sigma}(w,\bar{t}^{\prime}) > +
< j_{\sigma}(w,\bar{t}^{\prime}) \eta_{\xi}(v,t) > + $$

$$\left.+ < j_{\sigma}(w,t)
\eta_{\xi}(v,\bar{t}^{\prime}) > \Biggr ] + 2 \delta_{\xi \sigma}
\delta(v-w) \right \} \cdot
$$

$$
\cdot \left \{\Biggl [ \int\limits_{x_0}^{y} dz_{\zeta}
\alpha(z,y)\Biggl [< F_{\zeta \nu}(z,t) j_{\mu}(u,t^{\prime}) >+
< F_{\zeta \nu}(z,t) \eta_{\mu}(u,t^{\prime}) > \Biggr ] \right. -
$$

$$
- \int\limits_{0}^{t} dt^{\prime \prime} \Biggl [ < j_{\nu}(y,t^{\prime})
j_{\mu}(u,t^{\prime \prime}) > + < j_{\nu}(y,t^{\prime})
\eta_{\mu}(u,t^{\prime \prime}) > + $$

$$
+ \left.< j_{\mu}(u,t^{\prime \prime}) \eta_{\nu}(y,t^{\prime}) > + 2
\delta_{\mu \nu} \delta(y-u) \delta(t^{\prime \prime}-t^{\prime}) \Biggr ]
 \right \}= $$

\be
=\frac{\partial}{\partial u_{\lambda}} \Biggl [ < F_{\lambda \mu}(u,t)
j_{\nu}(y,t^{\prime}) > + < F_{\lambda \mu}(u,t)
\eta_{\nu}(y,t^{\prime}) > \Biggr ],
\ee
where the terms with space-time derivatives are put to the
right hand sides. In derivation of equations (21) and (22) we used
the following formula [9]:

\be
< e^A B > = < e^A > \Biggl( < B > + \sum_{n=1}^{\infty} \frac{1}{n!} \ll A^n
B \gg \Biggr),
\ee
where $A$ and $B$ are two arbitrary operators.

The system of equations (11), (12), (15)-(19), (21), (22) is a
complete system of equations of bilocal approximation for
gauge-invariant correlators
$< \varphi^*(\bar{x}, \bar{t}) \Phi(\bar{x}, x, \tau)
\varphi(x,t) >,~ < \eta^*(\bar{x}, t^{\prime}) \Phi(\bar{x},
x, \tau) \varphi(x,t) >, \\
 <\varphi^*(x,t) \Phi(x,\bar{x},\tau)
\eta(\bar{x}, t^{\prime}) >,
< F_{\mu \nu}(x,t) F_{\lambda \rho}
(\bar{x}, \bar{t}) >,
 < F_{\mu \nu}(x,t) j_{\lambda}(\bar{x}, \bar{t})
>, \\
 < F_{\mu \nu}(x,t) \eta_{\lambda}(\bar{x}, \bar{t}) >,  <
j_{\mu}(x,t) j_{\nu}(\bar{x}, \bar{t}) > , <j_{\mu}(x,t) \eta_{\nu}(\bar{x},
\bar{t})>.$ To obtain equations of higher approximations, one should exploit
generating functionals (10) and (20) in the same way, as it was done for
bilocal approximation.

Hence, there are two alternative ways of investigation of higher
correlators. E.g., for threelocal correlators, the first way is to derive,
using the genera-\\
ting functionals (10) and (20), an exact system of
equations, where threelocal and bilocal cumulants are considered on the same
footing. The other way (based on the additional assumption, that bilocal
cumulants are dominant) is to obtain bilocal cumulants from the equations
(11), (12), (15)-(19),
(21), (22) and then to substitute them into the
system of equations for threelocal cumulants.

Note, that ANO strings enter
equations of bilocal approximation only through an additional term in
equation (17).  Without writing explicitly twenty equations of threelocal
approximation for correlators $< \varphi^* \Phi \varphi >,\\
 < \eta^* \Phi
\varphi >,~ <\varphi^* \Phi \eta >,~ < F_{\mu \nu} F_{\lambda \rho} >,~
<F_{\mu \nu} \eta_{\lambda} >,< F_{\mu \nu} j_{\lambda} > ,~ < j_{\mu}
j_{\nu} >, \\
 <j_{\mu}\eta_{\nu}>,~ < \varphi^* \Phi \varphi \eta_{\mu} >, ~
< \eta^* \Phi \varphi \eta_{\mu} >, ~
 < \varphi^* \Phi \eta \eta_{\mu}>, ~ <
\varphi^* \Phi \varphi j_{\mu} >, \\
 < \eta^* \Phi \varphi j_{\mu} >, <
\varphi^* \Phi \eta j_{\mu} >,  < F_{\mu \nu} F_{\lambda \rho} j_{\sigma} >
,  <F_{\mu \nu} F_{\lambda \rho} \eta _{\sigma} >,
 < F_{\mu \nu}
F_{\lambda \rho} F_{\sigma \xi} > , < j_{\mu} j_{\nu} j_{\lambda} > ,  <
j_{\mu} j_{\nu} \eta_{\lambda} >,  < j_{\mu}\eta_{\nu}\eta_{\lambda} >$,
one can see, that in the correlators  $< F_{\mu \nu} F_{\lambda \rho}
j_{\sigma} > ,~ < F_{\mu \nu} F_{\lambda \rho} \eta_{\sigma} >$ and $<
F_{\mu \nu} F_{\lambda \rho} F_{\sigma \xi}>$ there are groups of terms,
consisting of bilocal correlators, which have the coefficients, depending on
the ANO strings.  Therefore, ANO strings play nontrivial role in the
process of formation of the hierarchy of correlators, connecting correlators
of different orders.\\

\vspace{3mm}
\section{QCD with Spinless Quarks at $T = 0$ and $T > 0$}

\hspace{5mm} In this section we demonstrate, how one can apply the methods
of section 2 to derivation of equations for correlators in QCD with spinless
quarks, i.e. the quarks, for which it is possible to omit the term
$\frac{g}{2} \sigma_{\mu \nu} F_{\mu \nu}$ in the quadrized Dirac equation,
where $\sigma_{\mu \nu} \equiv \frac{i}{2} [\gamma _\mu, \gamma_{\nu}]$.
Then we generalize this approach to the case of finite temperatures.

The main problem in stochastic quantization of fermions is that the solution
of naive Langevin equation, obtained directly from the fermionic action, in
the physical limit, $t \rightarrow + \infty$, has an asymptotics $e^{-mt
+ ipx}$, i.e. badly defined in the chiral limit. To avoid this difficulty,
the stochastic process is modified by introducing a kernel into the Langevin
equation: $\frac{\delta S}{\delta \overline{\psi}(x,t)} \rightarrow \int dy
K(x,y) \frac{\delta S}{\delta \overline{\psi}(y,t)}$, where $K(x,y)$ should
be chosen in such a way as to precisely cancel the negative eigenvalues of
$\frac{\delta S}{\delta \overline{\psi}(x,t)}$. For example, in [14] $K(x,y)
= \left (i \frac{\hat{\partial}}{\partial x} + m \right )\delta(x-y)$, and
the asymptotics of the solution becomes $e^{-(p^2+m^2)t+ipx}$ as in the
bosonic case. Our goal is to choose such a kernel, that ensures explicit
gauge invariance of the right hand sides of equations for gauge-invariant
vacuum correlators, containing quarks, antiquarks and Grassmann stochastic
noises.

In order to get desirable asymptotics of the solutions at $t \rightarrow +
\infty$ (not $(-\infty)$), i.e. to deal, as usual, with the retarded Green
function of the Langevin equation, we shall use the Euclidean
$\gamma$-matrices, multiplied by $"i"$:

\be
\vec{\gamma} = \beta \vec{\alpha},~~ \gamma_4 = i \beta,~~
\mbox{where}~~~ \vec{\alpha} =
\left(
\begin{array}{ll}
0 & \vec{\sigma}\\
\vec{\sigma} & 0
\end{array}
\right), ~~ \beta =\left(
\begin{array}{ll}
1 & 0\\
0 & -1
\end{array}
\right),
\ee
which satisfy anticommutational relations $\{ \gamma_{\mu}, \gamma_{\nu}\} =
-2\delta_{\mu \nu}$.

Let us introduce the kernel $K(x,y,t) = (i \hat{D}(x,t) + m) \delta(x-y)$
into the naive Langevin equation, obtained from the Euclidean QCD action
with $\gamma$-matrices, defined in (24),

$$
S = \int dx \Biggl [\frac{1}{4} F^a_{\mu \nu} F^a_{\mu \nu} -
\bar{\psi}(i \hat{D} - m) \psi \Biggr ],~~~ \mbox{where}~~~ D_{\mu} =
\partial_{\mu} - ig A_{\mu}.
$$
For spinless quarks one gets the following
Langevin equations:

\be
\dot{\psi} = (D^2 - m^2) \psi + \theta,
\ee
\be
\dot{\bar{\psi}} = ({\cal{D}}^2 - m^2) \bar{\psi} + \bar{\theta},
\ee
where $\bar{\psi} \equiv \psi^+ \gamma_4, {\cal{D}}_{\mu} \equiv
\partial_{\mu} + ig A_{\mu}$.
The correlator of Grassmann noises changes correspondingly:
\be
< \theta_{\alpha}(x,t) \bar{\theta}_{\beta}(y,t^{\prime}) >_{\theta
\bar{\theta}}~ = 2 [ i \hat{D}(x,t) \delta(x-y) + m \delta(x-y) ]_{\alpha
\beta} ~ \delta(t - t^{\prime}),
\ee
where for an arbitrary functional of
the $\theta^{\prime}s$ we have the stochastic expectation values:

 \be
<F[\theta, \bar{\theta} ]>_{\theta \bar{\theta}}~ =~ \frac{\int~D\theta
D\bar{\theta} F [\theta, \bar{\theta} ] exp [-\frac{1}{2} \int dx dt
\bar{\theta}(x,t) (i \hat{D} + m)^{-1} \theta(x,t) ]}{\int D \theta
D\bar{\theta} exp [ -\frac{1}{2} \int dx dt \bar{\theta}(x,t) (i \hat{D} +
m)^{-1} \theta(x,t) ]}.
\ee

Solving (25) and (26) via Feynman-Schwinger path integral representation
with the initial conditions $\psi(x,0) = \bar{\psi}(x,0) = 0$ and using (27)
and (23), one obtains in bilocal approximation:
$$
 <\bar{\psi}_{\beta}(\bar{x}, \bar{t}) \Phi(\bar{x}, x, \tau)
\psi_{\alpha}(x,t) > =
$$

$$
= 2\int\limits_{0}^{min(t,\bar{t})} dt^{\prime}~\int d \bar{y} (D
\bar{z})_{\bar{x} \bar{y}} K_{\bar{z}}(\bar{t}, t^{\prime}) \int \left \{ -i
\Biggl [\frac{\hat{\partial}}{\partial y}(Dz)_{xy} K_z(t, t^{\prime}) \Biggr
]_{y=\bar{y}} \right. + $$

\be
\left. +m(Dz)_{xy}K_z(t, t^{\prime})_{\mid_{y=\bar{y}}} \right
\}_{\alpha\beta} < W_{\bar{y} \bar{x} x \bar{y}} > ,
\ee
where $< ...>$ means the
averaging over the vacua of $\eta^a_{\mu}$ and $(\theta, \bar{\theta}),
\eta^a_{\mu}$ is a Gaussian stochastic noise with the bilocal correlator $<
\eta^a_{\mu}(x,t) \eta^b_{\nu}(y,t^{\prime}) >_{\eta} =$\\
$=2 \delta_{\mu
\nu} \delta^{ab} \delta(x-y) \delta(t-t^{\prime})$.  Here

$$
\Phi(\bar{x}, x, \tau)= P~ exp \Biggl [ ig \int\limits_{x}^{\bar{x}} dw_{\mu}
A_{\mu}(w, \tau) \Biggr ],
$$

\be
W_{\bar{y}\bar{x} x \bar{y}} =P exp \left \{ ig
\Biggl [\int\limits_{\bar{x}}^{\bar{y}} d\bar{z}_{\mu} A_{\mu}
(\bar{z}(\bar{\xi}), \bar{\xi}) + \int\limits_{x}^{\bar{x}} dw_{\mu}
A_{\mu}(w,\tau) + \int\limits_{\bar{y}}^{x} dz_{\mu} A_{\mu}(z (\xi),
\xi)\Biggl ] \right \}.
\ee

To perform differentiation in the equation (29), let us extract explicitly
the dependence on the point $y$ from the measure of integration and from
the kinematical factor $K_z(t, t^{\prime})$. To this end we pass to the
integration over trajectories
\be
u(\xi) = z(\xi) +
\frac{y-x}{t-t^{\prime}}(\xi-t^{\prime}) - y.
\ee
After that differentiation
may be easily performed, and the result has the form

$$
< \bar{\psi}_{\beta}(\bar{x}, \bar{t})\Phi(\bar{x}, x, \tau)
\psi_{\alpha}(x,t) > =
$$

$$
= 2\int\limits_{0}^{min(t,\bar{t})} dt^{\prime}~\int d \bar{y}(D\bar{z})
_{\bar{x} \bar{y}}K_{\bar{z}}(\bar{t},t^{\prime})\Biggl [\frac{i}{2}
\frac{\hat{x}-\hat{\bar{y}}}{t^{\prime} - t}
e^{\frac{(x-\bar{y})^2}{4(t^{\prime}-t)}}
 ~\int (Du)_{00} \cdot
$$

\be
\cdot K_u(t, t^{\prime}) < W_{\bar{y} \bar{x} x \bar{y}} > + m \int (Dz)_{xy}
K_z(t, t^{\prime})_{y=\bar{y}} < W_{\bar{y}\bar{x} x \bar{y}} >
\Biggr ]_{\alpha \beta}.
\ee
Here in the first Wilson loop in the right hand
side it is implied, that in the last term in (30) $z_{\mu}$ is expressed
through $u_{\mu}$ via (31).

In analogous way one may obtain equations for two other gauge-invariant
correlators, containing spinor fields, in bilocal approximation:

\be
< \bar{\psi}_{\beta}(\bar{x}, \bar{t}) \Phi(\bar{x}, x, \tau)
\theta_{\alpha}(x,t) >~ \mbox{and}~ < \bar{\theta}_{\beta}(\bar{x}, \bar{t})
\Phi (\bar{x}, x, \tau) \psi_{\alpha}(x,t) >.
\ee
In contrast to the AHM, in the case of QCD it is, of course, impossible to
calculate exact Green function of the Langevin equation for $F^a_{\mu
\nu}(x,t)$ (this problem is equivalent to the solution of Yang-Mills
equation in the general form). Therefore, in order to obtain all the rest
equations for gauge-invariant correlators in bilocal approximation, one
should act in the same way as it was done in section 2 for derivation of
equations (21) and (22). Using Schwinger gauge and introducing the
generating functional
\be
\Phi_{\beta} = P exp \left \{ ig \oint\limits_{C} dx_{\mu} \Biggl [
\int\limits_{x_0}^{x} dz_{\nu} \alpha(z,x) F_{\nu \mu}(z,t) + \beta
\int\limits_{0}^{t} dt^{\prime} (gj_{\mu}(x,t^{\prime}) +
\eta_{\mu}(x,t^{\prime})) \Biggr ] \right \},
\ee
where $j^a_{\mu} \equiv
\bar{\psi}\gamma_{\mu} t^a \psi$ is a quark current, differentiating (34)
by $t$ and applying to it cumulant expansion, one obtains, due to the
Langevin equation

$$
\dot{A}^a_{\mu} = (D_{\lambda} F_{\lambda \mu})^a + g j^a_{\mu} +
\eta^a_{\mu},
$$
where $D_{\lambda}$ is the ajoint covariant derivative, the first equation.
After that one should differentiate it five times by $\beta$, use the formula
(23) and in all the equations put $\beta$ equal to $(-1)$. Finally, to
obtain a minimal closed system of equations of bilocal approximation, one
should avoid $D_{\lambda}$ for $\partial_{\lambda}$ in the right hand sides
of these equations. This may be done, using the expressions for higher
orders$^{\prime}$ path-ordered cumulants [3,10] and the formula

$$
tr < \Phi(x_0, u,t) (D_{\lambda} F_{\lambda \mu}
(u,t^{\prime}))\Phi(u,x_0,t) G_{\mu_1...\mu_n} > = $$

$$= tr \left \{\frac{\partial}{\partial u_{\lambda}} <F_{\lambda
\mu}(u,x_0,t)G_{\mu_1...\mu_n} > +
 ig \int\limits_{x_0}^{u} dx_{\rho}
\alpha(x,u) \cdot \right.$$

$$\cdot \Biggl [ < F_{\lambda \mu}(u,x_0,t) G_{\mu_1...\mu_n}
F_{\lambda \rho}(x,x_0,t) > -
$$

$$ \left.- < F_{\lambda \mu}(u,x_0,t) F_{\lambda
\rho}(x,x_0,t)G_{\mu_1...\mu_n} > \Biggr] \right \},
$$
where $F_{\mu \nu}(x,x_0,t) \equiv \Phi(x_0,x,t) F_{\mu
\nu}(x,t) \Phi (x,x_0,t)$, any parallel transporter between
$x_0$ and any other point goes along straight line,
$G_{\mu_1...\mu_ n} \equiv $\\
 $\equiv G_{\mu_1...\mu_ n}(x_1, t_1,..., x_n,
t_n, x_0, t)$ is, generally speaking, a product of some number of
$F_{\mu \nu}, j_{\mu}$ and $\eta_{\mu}$, which are given in the points
$x_1,...,x_n$ at the moments $t_1,...,t_n$ of Langevin time
respectively, and all the parallel transporters \\
between $x_0$ and
each of these points are given at the same moment $t$. This
method of derivation of the system of equations in the case of
pure gluodynamics was used in [10].

After that, substituting $<F_{\mu \nu}(x,x_0,t) F_{\lambda
\rho}(y,x_0,t) >$, obtained from this system of six equations for
correlators
$$
< F_{\mu \nu}(x, x_0, t) F_{\lambda \rho}(y, x_0, t) >,~ <
F_{\mu \nu}(x, x_0, t) j_{\lambda}(y, x_0, t, t^{\prime}) >,
$$

$$< F_{\mu \nu}(x, x_0, t)\eta_{\lambda}(y, x_0, t, t^{\prime}) >,~
<j_{\mu}(x, x_0, t, t^{\prime})j_{\nu}(y, x_0, t, t^{\prime \prime}) >,
$$

\be
< j_{\mu}(x,x_0,t,t^{\prime})
\eta_{\nu}(y, x_0, t, t^{\prime \prime}) > ,~ < \eta_{\mu}(x, x_0, t,
t^{\prime})  \eta_{\nu}(y, x_0, t, t^{\prime \prime}) >,
\ee
where
$$
j_{\mu}(x,
x_0, t, t^{\prime}) \equiv \Phi(x_0, x,t) j_{\mu}(x,t^{\prime}) \Phi(x,
x_0, t),
$$

$$
\eta_{\mu}(x, x_0, t, t^{\prime}) \equiv \Phi(x_0, x, t)
\eta_{\mu}(x,t^{\prime})\Phi(x, x_0, t),
$$
into the averaged Wilson loops, expanded up to the second cumulant
via nonabelian Stokes theorem in (32) and in the right hand sides of
equations for correlators (33), we obtain all the correlators of bilocal
approximation.

In order to obtain equations for higher correlators, one should
introduce the analog of generating functional (10) and follow
the algorithm, described in section 2.

To conclude this section, we generalize the  method of derivation of
the system of exact equations, suggested above, to the case of
QCD at finite temperatures. It was shown in [15], that  the ordinary
results of finite temperature field theory may be reproduced in the
framework of stochastic quantization, if one requires the stochastic noise
to be periodic for bosons (antiperiodic for fermions) in the coordinate
$x_4$ with the period $\beta \equiv \frac{1}{T}$. Therefore, we supply
equations (25) and (26) with the following conditions:
\be
\theta(\vec{x}, x_4, t) = -\theta(\vec{x}, x_4 + \beta, t), ~~
\bar{\theta}(\vec{x}, x_4, t) =
-\bar{\theta}(\vec{x}, x_4 + \beta, t)
\ee
and modify the meaning of the averaging over $(\theta,
\bar{\theta})$ from (28) to
$$
< F [\theta, \bar{\theta} ] >_{\theta \bar{\theta}} = \frac {\int D \theta
D\bar{\theta} F [\theta, \bar{\theta} ] exp \Biggl[-\frac{1}{2}
\int\limits_{0}^{\beta} dx_4 \int d\vec{x} dt \theta(x,t)(i
\hat{D}+m)^{-1}\theta(x,t) \Biggr]}{\int D\theta D\bar{\theta}exp \Biggl [
-\frac{1}{2} \int\limits_{0}^{\beta} dx_4 \int d \vec{x}dt
\bar{\theta}(x,t)(i \hat{D} + m)^{-1} \theta(x,t) \Biggr]}.
$$
Then it is
easy to check, that in the absence of gluonic fields, the solution of
equation (25) with initial condition $\psi(x,0) = 0$ has the form
\be
\psi(x,t) = \int\limits_{0}^{\infty} dt^{\prime}\int\limits_{0}^{\beta}dy_4
\int d\vec{y}G(x-y, t-t^{\prime})\theta(y,t^{\prime}),
\ee
where
$$
G(x,t)
\equiv \frac{\theta(t)}{\beta}\sum\limits_{n=-\infty}^{+ \infty} \int
\frac{d\vec{k}}{(2\pi)^3}e^{ik_nx -(k^2_n+m^2)t},
$$

$$
k_n \equiv (\vec{k}, \omega^-_n),~~ \omega^-_n \equiv \frac{(2n
+1)\pi}{\beta}. $$
Solving equation (26) for a free antiquark with the
initial condition $\bar{\psi}(x,0) =$\\
$=0$ in analogous way, one gets after
some calculations

$$
< \bar{\psi}_{\beta}(x^{\prime}, t^{\prime})\psi_{\alpha}(x,t) > _{\theta
\bar{\theta}} = \frac{1}{\beta}\sum\limits_{n=-\infty}^{+ \infty} \int
\frac{d\vec{k}}{(2\pi)^3}(-\hat{k}_n + m)_{\alpha
\beta}e^{ik_n(x-x^{\prime})}\cdot
$$

$$
\cdot \frac {e^{-(k^2_n+m^2)\mid t-t^{\prime} \mid}-
e^{-(k^2_n+m^2)(t+t^{\prime})}} {k^2_n + m^2}, $$
that in the physical limit
$t = t^{\prime}\rightarrow + \infty$ yields standard result for the
propagator of a free fermion at finite temperature [16].

In the case, when gluonic fields are present, one, solving equation (25) with
the initial condition $\psi(x,0)=0$ through Feynman-Schwinger path integral
representation, obtains instead of (37):

$$
\psi(x,t) = \int\limits _{0}^{t}dt^{\prime}\int dy (Dz)^w_{xy}K_z(t,
t^{\prime}) \Phi(x,y)\theta(y,t),
$$
where

$$
(Dz)^w_{xy}\equiv \lim_{N \rightarrow \infty} \prod\limits_{l=1}^{N}\frac{d^4
\zeta(l)}{(4\pi \epsilon)^2} \sum_{n=- \infty}^{+
\infty}\frac{d^4 p}{(2\pi)^4
}e^{ip_{\mu}\Biggl (\sum\limits_{k=1}^{N}\zeta_{\mu}(k)-(x-y)_{\mu}-n \beta
\delta_{\mu 4}\Biggr )},
$$

$$
\zeta_{\mu}(k) \equiv z_{\mu}(\epsilon k) -
z_{\mu}(\epsilon(k-1))~~[17],
$$

$$
\Phi(x,y)\equiv P exp \Biggl [ig~ \int\limits_{y}^{x}dz_{\mu}
A_{\mu} (z(\xi), \xi)\Biggr].
$$
Therefore, one can check, that, for example,
equation (32) changes as follows:
$$
<\bar{\psi}_{\beta}(\bar{x},\bar{t})\Phi(\bar{x},x,\tau)\psi_{\alpha}(x,t)>
=2\int\limits_{0}^{min (t,\bar{t})}dt^{\prime}\int
d\bar{y}(D\bar{z})^w_{\bar{x}\bar{y}}K_{\bar{z}}(\bar{t},t^{\prime})
\Biggl
[\frac{i}{2}\frac{\hat{x}-\hat{\bar{y}}}{t^{\prime}-t}
e^{\frac{(x-\bar{y})^2}{4(t^{\prime}-t)}}\cdot
$$

\be
\cdot \int\limits_{u(t^{\prime})=u(t)=0}(Du)^w K_u(t,t^{\prime})<
W_{\bar{y}\bar{x}x\bar{y}}>+m
\int(Dz)^w_{xy}K_{z}(t,t^{\prime})_{\mid_{y=\bar{y}}}<
W_{\bar{y}\bar{x}x \bar{y}}> \Biggr ]_{\alpha \beta},
\ee
where

\be
(Du)^w = \lim\limits_{N \rightarrow
\infty}\prod\limits_{l=1}^{N}\frac{d^4\lambda(l)}{(4\pi \epsilon)^2}
\sum_{n=-\infty}^{+\infty}\frac{d^4p}{(2\pi)^4}e^{ip_{\mu}
\Biggl (\sum\limits_{k=1}^{N}\lambda_{\mu}(k)-n \beta
\delta_{\mu 4}\Biggr )},
\ee

$$
\lambda(k) \equiv u(\epsilon k) - u(\epsilon(k-1)).$$
Solving equations for correlators (35), one should keep in mind, that the
stochastic noise fields $\eta_{\mu}^a(x,t)$ are periodic in the coordinate
$x_4$ with the period $\beta$.

\vspace{3mm}
\section{QED: Including Spin Effects}

\hspace{5mm} In this section we shall demonstrate, how one can exactly take
into account spin effects in the framework of suggested approach. Let us
consider for simplicity the case of QED and, instead of equation (25), solve
exact Langevin equation

\be \dot{\psi}= -(\hat{D}^2+m^2)\psi + \theta,
\ee
where
$$
\hat{D}^2 = -(D^2 + \frac{e}{2} \sigma_{\mu \nu}F_{\mu \nu}),
$$
via Feynman-Schwinger path integral representation. We shall follow the
method, suggested in [11,12] for derivation of a propagator of a spinning
particle in the external Abelian field. The retarded Green function of
equation (40) has the form:
$$
G(x,y,t)=\theta(t)\int (Dz)_{xy}Dp~ exp  \left \{i \int\limits_{0}^{t} d\tau
\Biggl [p\dot{z}+i(p_{\mu}-e A_{\mu}(z,\tau))^2 +\right.
$$

$$
\left.+\frac{e}{2}F_{\mu \nu}(z,\tau) \frac{\delta^2}{\delta \rho_{\mu}\delta
\rho_{\nu}} + im^2 \Biggr ] \right \} T exp \Biggl[ \int\limits_{0}^{t}
\rho_{\lambda}(\tau^{\prime})\gamma_{\lambda} d\tau^{\prime} \Biggl
]_{\mid_{\rho=0}}, $$
where  $\rho_{\mu}$ are Grassmann sources,
anticommuting with $\gamma$--matrices. Using Wick theorem [18], we get:

$$ T exp \Biggl
[\int\limits_{0}^{t}\rho_{\lambda}(\tau^{\prime})\gamma_{\lambda}d
\tau^{\prime}\Biggr ] = $$

$$
=exp \Biggl [-\frac{1}{2}\int\limits_{0}^{t}d\tau_1 \int\limits_{0}^{t}d
\tau_2 \rho_{\mu}(\tau_1) sgn(\tau_1-\tau_2) \rho_{\mu}(\tau_2) \Biggr ] exp
\Biggl[ \int\limits_{0}^{t}\rho_{\lambda}(\tau)\gamma_{\lambda}d \tau \Biggr
] =
$$

$$
= \frac{\int\limits_{\xi(0)+\xi(t)=0} D\xi~exp \left \{
\int\limits_{0}^{t} d\tau \Biggl [
\frac{1}{4}\xi_{\mu}(\tau)\dot{\xi}_{\mu}(\tau)-
i\rho_{\mu}(\tau)\xi_{\mu}(\tau)\Biggr] \right \}}{\int~D\xi~ exp
\Biggl [\frac{1}{4} \int\limits_{0}^{t} d\tau
\xi_{\mu}(\tau)\dot{\xi}_{\mu}(\tau) \Biggr ]} \cdot
$$

$$ \cdot exp \Biggl(i
\gamma_{\mu}\frac{\partial}{\partial \theta_{\mu}} \Biggr )exp \Biggl [
-i\int\limits^t_{0} \rho_{\mu}(\tau)\theta_{\mu}d \tau
\Biggr ]_{\mid_{\theta=0}},
$$
where $\xi_{\mu}(\tau)$ are odd trajectories.
To simplify this expression, one can pass to the integration over
trajectories $\chi(\tau)= \frac{1}{2}(\xi(\tau)+ \theta)$ and obtain:

$$
G(x,y,t)= \theta(t)\int (Dz)_{xy} exp \Biggl(i
\gamma_{\mu}\frac{\partial}{\partial \theta_{\mu}} \Biggr)
\int\limits_{\chi_{\mu}(0)+\chi_{\mu}(t)=\theta_{\mu}} D\chi~exp \left \{
\int\limits^{t}_{0} d\tau \Biggl[-\frac{\dot{z}^2}{4} - m^2 + \right.
$$

$$
\left.+ ie A_{\mu}
\dot{z}_{\mu}- 2 ie F_{\mu \nu} \chi_{\mu}\chi_{\nu}+
\chi_{\mu}\dot{\chi}_{\mu}\Biggr] + \chi_{\mu}(t)\chi_{\mu}(0)
\right \}_{\mid_{\theta_{\mu}=0}}.
$$
In order to avoid the restriction to
the trajectories, over which we perform the integration, let us pass to the
integration over trajectories $\omega(\tau)$ , such that $\chi(\tau)=
\frac{1}{2}\int\limits^{t}_{0}sgn
(\tau-\tau^{\prime})\omega(\tau^{\prime})d\tau^{\prime}+
\frac{1}{2}\theta$. Noticing, that for an arbitrary function, given at the
Grassmann algebra, the following formula takes place:

$$
exp \Biggl(i \gamma_{\mu}\frac{\partial}{\partial \theta_{\mu}} \Biggr )
f(\theta)\mid_{\theta=0} = f \Biggl(\frac{\partial}{\partial \zeta}
\Biggr)exp (i \zeta_{\mu}\gamma_{\mu})_{\mid_{\zeta=0}},
$$
one obtains
the following expression for $G(x,y,t)$:
$$ G(x,y,t) =
\frac{1}{2}\theta(t)\int(Dz)_{xy}exp \left \{\int\limits^{t}_{0}d\tau \Biggl
[ -\frac{\dot{z}^2}{4} - m^2 +ie A_{\mu}(z, \tau)\dot{z}_{\mu} - \right.$$

$$ \left. - \frac{ie}{2} F_{\mu\nu}(z,\tau)\frac{\partial^2}{\partial
\zeta_{\mu}\partial \zeta_{\nu}} \Biggr ] \right \}
\int D\omega~ exp \left \{\int\limits^{t}_{0} d\tau \Biggl[-
\frac{ie}{2}\int\limits^{t}_{0} d\tau^{\prime} \cdot \right.$$

$$\cdot sgn(\tau-\tau^{\prime})
\omega_{\mu}(\tau^{\prime})F_{\mu \nu}(z,\tau)
\int\limits^{t}_{0} d \tau^{\prime \prime} sgn
(\tau-\tau^{\prime \prime})\omega_{\nu}(\tau^{\prime \prime}) +
$$

$$
+ \frac{1}{2}\int\limits^{t}_{0}
d\tau^{\prime}sgn(\tau-\tau^{\prime})\omega_{\mu}(\tau^{\prime})
\omega_{\mu} (\tau) - \frac{ie}{2} \int\limits^{t}_{0} d\tau^{\prime}
sgn(\tau-\tau^{\prime})\omega_{\mu}(\tau^{\prime})F_{\mu
\nu}(z,\tau)\frac{\partial}{\partial \zeta_{\nu}}-
$$

$$
-\frac{ie}{2}\frac{\partial}{\partial
\zeta_{\mu}}F_{\mu \nu}(z,\tau) \int\limits^{t}_{0} d\tau^{\prime \prime}
sgn(\tau-\tau^{\prime \prime})\omega_{\nu}(\tau^{\prime \prime}) +\frac{1}{2}
\frac{\partial}{\partial \zeta_{\mu}} \omega_{\mu}(\tau) \Biggr] -$$

\be
\left.-\frac{1}{4}\int\limits^{t}_{0} d\tau^{\prime}\int\limits^{t}_{0}d
\tau^{\prime \prime} \omega_{\mu}(\tau^{\prime})
 sgn
(\tau^{\prime}-\tau^{\prime \prime})\omega_{\mu}(\tau^{\prime \prime})
\right \} exp (i \zeta_{\mu}\gamma_{\mu})_{\mid_{\zeta=0}}.
\ee

$\int D\omega {...}$ may be represented in the form
$$
\int D\omega~ exp \Biggl [
-\frac{1}{2}\omega_{\mu}T_{\mu \nu} (z,e)\omega_{\nu}+ I_{\mu}\omega_{\mu}
\Biggr ],
$$
where

$$
T_{\mu \nu}(z,e)=ie \epsilon F_{\mu \nu}(z){\epsilon} +
\frac{3}{2}\delta_{\mu \nu}\epsilon,~
I_{\mu}= ie F_{\mu \nu}(z)\epsilon \frac{\partial}{\partial \zeta_{\nu}}+
\frac{1}{2}\frac{\partial}{\partial \zeta_{\mu}},
$$
and we used the following notation: for example,
$$\omega \epsilon \omega
\equiv
\int\limits^{t}_{0}d\tau^{\prime}\int\limits^{t}_{0}
d\tau^{\prime \prime}\omega(\tau^{\prime}) sgn
(\tau^{\prime}-\tau^{\prime \prime})\omega(\tau^{\prime \prime}).
$$
Here and everywhere
later, where such a notation is used, the Langevin time argument of $F_{\mu
\nu}$ is omitted, since it is integrated over. Then
$$ \int
D\omega{...}= \left [\frac{det T(z,e)}{det T(z,0)}\right ]^{1/2} exp \left
\{ -\frac{1}{2} I_{\mu}[ T^{-1}(z,e)]_{\mu \nu}~I_{\nu } \right \}=
$$
$$
= exp \left \{ i \int\limits^{e}_{0}de^{\prime} tr  [ Q_{\mu
\nu}(z,e^{\prime})F_{\mu \nu}(z) ] \right \} exp
\Biggl [-\frac{1}{2}I_{\mu}R_{\mu \nu}(z,e)I_{\nu} \Biggr],
$$
where
$$
Q_{\mu
\nu}(z,e) \equiv  \epsilon [ T^{-1}(z,e)]_{\mu \nu}\epsilon,~~ R_{\mu
\nu}(z,e)\equiv [T^{-1}(z,e)] _{\mu \nu}.
$$

Differentiation in (41) with the help of the formula
$$
exp \left (B_{\mu \nu}\frac{\partial^2}{\partial \zeta_{\mu}\partial
\zeta_{\nu}}\right )exp (i \zeta_{\lambda}
\gamma_{\lambda})_{\mid_{\zeta=0}} = 1 - i \sigma_{\mu \nu} B_{\mu \nu}+
i \gamma_5 B_{\mu \nu} B^*_{\mu \nu},
$$
where
$$
\gamma_5 \equiv -i\gamma_1 \gamma_2 \gamma_3 \gamma_4, ~~~
"^*" \mbox{denotes the dual tensor},
$$
yields the final result:
$$
G(x,y,t)= \frac{\theta(t)}{2} exp \left \{\frac{1}{2}\Biggl [e
(e-2i)-\frac{1}{4}\Biggr ] \right \} \int (Dz)_{xy} exp
\left \{\int\limits^{t}_{0} d\tau \Biggl [-\frac{\dot{z}^2}{4}-m^2 +
\right. $$

$$ + ie
A_{\mu}(z,\tau)\dot{z}_{\mu} -
 \frac{e}{2}\sigma_{\mu \nu}
F_{\mu \nu}(z,\tau) + \frac{e}{2} \gamma_5 \int\limits^{t}_{0}
d\tau^{\prime} F_{\mu \nu}(z,\tau)F^*_{\mu \nu}(z,\tau^{\prime}) \Biggr
] + $$

$$ \left.+ i \int\limits^{e}_{0} de^{\prime} tr [Q_{\mu
\nu}(z,e^{\prime})F_{\mu \nu}(z) ] + \Lambda(z,e) \right \}, $$
where 

 $$ \Lambda(z,e) =
-\frac{i}{2} e^2 \Biggl [\sigma_{\lambda \rho} F_{\mu \lambda}(z) \epsilon
R_{\mu \nu}(z,e)F_{\nu \rho}(z) \epsilon - \gamma_5(F_{\mu \lambda}(z)
\epsilon R_{\mu \nu}(z,e) F_{\nu \rho}(z) \epsilon) \cdot $$

$$ \cdot (F_{\alpha \lambda}(z) \epsilon R_{\alpha \beta}(z,e)
F_{\beta \rho} (z)\epsilon)^* \Biggr ] -\frac{e}{4} \Biggl [\sigma_{\lambda
\nu} F_{\mu \lambda}(z) \epsilon R_{\mu \nu}(z,e) + \sigma_{\mu \rho} R_{\mu
\nu}(z,e) F_{\nu \rho}(z) \epsilon - $$

$$ - \gamma_5(F_{\mu \lambda}(z)
\epsilon R_{\mu \nu}(z,e))(F_{\alpha \lambda}(z) \epsilon R_{\alpha
\nu}(z,e))^* - \gamma_5 (R_{\mu \nu}(z,e) F_{\nu \rho}(z) \epsilon)\cdot$$

$$\cdot (R_{\mu
\alpha}(z,e) F_{\alpha \rho}(z) \epsilon)^* \Biggr ] +
\frac{i}{8}\Biggl [\sigma_{\mu \nu} R_{\mu \nu}(z,e) - \gamma_5 R_{\mu
\nu}(z,e) R^*_{\mu \nu}(z,e)\Biggr ]. $$

In the same way one may obtain the Green function of the Langevin equation
$$
\dot{\bar{\psi}} = -\bar{\psi}\stackrel{\longleftarrow}{(\hat{\cal{D}}^2 +
m^2)} + \bar{\theta}, $$
where

$$ \hat{\cal{D}}^2 = -{\cal{D}}^2 +
\frac{e}{2} \sigma_{\mu \nu} F_{\mu \nu}. $$

To generalize the equation (32) to the spinor case, one should again, for the
group of terms in the functional integral, pass to the integration over
trajectories (31). The result has the form:
$$
<\bar{\psi}_{\beta}(\bar{x}, \bar{t})\Phi(\bar{x}, x, \tau)
\psi_{\alpha}(x,t) >= 2\int\limits^{min(t,\bar{t})}_{0}dt^{\prime} \int
d\bar{y}(D\bar{z})_{\bar{x}\bar{y}} K_{\bar{z}}(\bar{t}, t^{\prime},-e)
\cdot $$

$$\cdot \left \{\frac{i}{2(t^{\prime}-t)}(\hat{x}-\hat{\bar{y}})_{\gamma
\delta} e^{\frac{(x-\bar{y})^2}{4(t^{\prime}-t)}}
 \int(Du)_{00}K_u(t,t^{\prime},e) \cdot \right.$$

$$ \cdot < W_{\bar{y}\bar{x}x\bar{y}} \left
\{ exp \left \{\frac{e}{2}\int\limits^{t}_{t^{\prime}} d\xi \Biggl
[-\sigma_{\mu \nu}F_{\mu \nu}(z,\xi)+\gamma_5 \int\limits^{t}_{t^{\prime}}
d\xi^{\prime}F_{\mu \nu}(z,\xi)F^*_{\mu \nu}(z,\xi^{\prime}) \Biggr]
+ \right. \right.$$

$$ + \left. \left. i\int\limits^{e}_{0} de^{\prime} tr[Q_{\mu
\nu}(z,e^{\prime})F_{\mu \nu}(z)] + \Lambda(z,e)
\right \} \right\}_{\alpha
\gamma\mid_{z=\frac{x-\bar{y}}
{t-t^{\prime}}(\xi-t^{\prime})+u(\xi)+y}}
\left \{exp \left \{ \frac{e}{2}\int\limits^{\bar{t}}_{t^{\prime}} d\bar{\xi}
\Biggl [\sigma_{\lambda \rho} \cdot \right. \right.
$$

$$
\cdot F_{\lambda \rho}(\bar{z},
\bar{\xi})-\gamma_5 \int\limits^{\bar{t}}_{t^{\prime}} d\bar{\xi}^{\prime}
F_{\lambda \rho} (\bar{z}, \bar{\xi}) F^*_{\lambda \rho}
(\bar{z},\bar{\xi}^{\prime})\Biggr ] + $$

$$ \left.\left. + i \int\limits^{e}_{0} de^{\prime
\prime} tr [Q_{\lambda \rho}(\bar{z},e^{\prime \prime}) F_{\lambda
\rho}(\bar{z})] + \Lambda(\bar{z}, -e) \right \} \right \}_{\delta\beta}
> +
$$

$$ +m \int (Dz)_{x \bar{y}}
K_z(t,
t^{\prime},e)_{\mid_{y=\bar{y}}} < W_{\bar{y} \bar{x} x \bar{y}} \left \{exp
\left \{\frac{e}{2} \int\limits^{t}_{t^{\prime}} d\xi \Biggl [-\sigma_{\mu
\nu} F_{\mu \nu}(z,\xi) + \right.\right. $$

$$ \left. \left.+ \gamma_5 \int\limits^{t}_{t^{\prime}} d\xi^{\prime}
F_{\mu \nu}(z,\xi)
  F^*_{\mu \nu}(z,
\xi^{\prime})\Biggr ] + i \int\limits^{e}_{0} de^{\prime} tr [Q_{\mu
\nu}(z,e^{\prime}) F_{\mu \nu}(z) ] + \Lambda(z,e) \right \} \right
\}_{\alpha \gamma} \cdot$$

$$\cdot \left \{exp \left \{\frac{e}{2}
\int\limits^{\bar{t}}_{t^{\prime}} d\bar{\xi}\Biggl [\sigma_{\lambda \rho}
F_{\lambda \rho}(\bar{z}, \bar{\xi}) -
\gamma_5
\int\limits^{\bar{t}}_{t^{\prime}} d\bar{\xi}^{\prime} F_{\lambda
\rho}(\bar{z}, \bar{\xi}) F^*_{\lambda
\rho}(\bar{z},\bar{\xi}^{\prime}) \Biggr ] + \right. \right.$$

\be
\left.\left.\left. + i \int\limits^{e}_{0}
de^{\prime \prime} tr [Q_{\lambda \rho}(\bar{z},e^{\prime \prime})
F_{\lambda \rho}(\bar{z})] + \Lambda(\bar{z}, - e) \right \} \right
\}_{\gamma \beta} > \right\},
\ee
where

$$ K_z(t,t^{\prime},e) \equiv
\frac{1}{2} exp \Biggl[-\int\limits^t_{t^{\prime}} \frac{\dot{z}^2}{4} d\xi
- m^2 (t-t^{\prime}) + \frac{e^2}{2} - ie - \frac{1}{8}
\Biggr].  $$

Therefore, we reduced the right hand side of the equation for\\
$<\bar{\psi}_{\beta}(\bar{x}, \bar{t}) \Phi(\bar{x}, x, \tau)
\psi_{\alpha}(x,t)>$ to the path integral over bosonic trajectories from the
functional, whose spinor structure is given explicitly. The following steps
of derivation of the system of equations for gauge-invariant correlators are
similar to the previous two sections and based on the introduction of the
corresponding generating functionals, analogous to (10) and (20).

\vspace{3mm}
\section{Conclusion}

\hspace{5mm} In this paper, using stochastic quantization method [1], we developed a
systematical procedure of derivation of the systems of equations for
gauge-invariant vacuum correlators in different field theories. In section 2
on the  example of AHM we demonstrated, how this method may be applied to
quantization of classical solutions (ANO strings); the quantization of more
complicated objects will be treated elsewhere. The generating functionals
(10) and (20), introduced in this section, appropriately modified, may be
used to derivation of equations in any other gauge theory with the help of
the algorithm, described in sections 2 and 3. This algorithm, was used in
section 2, where the  minimal set of equations of bilocal
approximation  was derived, and the role of ANO strings in the higher
correlators was discussed. The suggested approach yields the gauge-invariant
method of quantization of the theories with spontaneous symmetry breaking;
this problem will be investigated in the future publications.

According to the general principa of the Method of Vacuum Correlators, the
obtained system of equations is splitted into two parts: the part, containing
pure gauge fields$^{\prime}$ correlators and matter fields$^{\prime}$
currents and the part, containing matter Green functions, so that the
equations of the first subsystem may be solved independently from the second
equations, but not vice versa. The connection of this fact with the
large--$N$ behaviour of QCD and a new type of the  Master field equation,
 based on the suggested approach, will be treated elsewhere.

 The method of derivation of equations in QCD with spinless quarks was
 generalized at the end of section 3 to the case of finite temperatures, and
 in section 4 we showed on the example of QED, how one can take  into
 account spin effects exactly.

The equations, obtained in the case of QCD, contain both perturbative and
nonperturbative gluonic contributions to the vacuum correlators. The problem
of separation of perturbative gluonic contributions in all  the terms of
cumulant expansion and the regularization of the obtained equations will
be the topics of  separate publications.

The results, presented in this paper, were partially reported at the
International Workshop "Nonperturbative Approaches to QCD", Trento, Italy,
July 10-29, 1995.

\vspace{3mm}
\section{Acknowledgements}

\hspace{5mm} I would like to thank Professor Yu.A.Simonov for useful
discussions,
Yu.S.Kalashnikova, who has told me about the papers [11] and M.Markina for
typing the manuscript.

The work is
supported by the Russian Fundamental Research Foundation, Grant
No.93-02-14937 and by the INTAS, Grant No.94--2851.

\newpage
\begin{flushright}
{\it Appendix}
\end{flushright}

{\Large\bf Derivation of the Green function (8)}

\vspace{5mm}
\setcounter{equation}{0}
\def\theequation{A.\arabic{equation}}

Below we present explicit derivation of a Green function of the Langevin
equation
\be
\dot{F}_{\mu \nu}(x,t) + \partial_{\nu} \partial_{\lambda} F_{\lambda
\mu}(x,t) - \partial_{\mu} \partial_{\lambda}F_{\lambda \nu}(x,t) = f_{\mu
\nu}(x,t),
\ee
where
$$
f_{\mu \nu}(x,t) = \partial_{\mu}(j_{\nu}(x,t) + \eta_{\nu}(x,t)) -
\partial_{\nu}(j_{\mu}(x,t) + \eta_{\mu}(x,t)).
$$
Writing equation (A.1) in the momentum representation:
\be
\dot{F}_{\mu \nu}(k,t) + k_{\mu}k_{\lambda} F_{\lambda
\nu}(k,t) + k_{\nu}k_{\lambda}F_{\mu \lambda}(k,t) = f_{\mu
\nu}(k,t),
\ee
one can check, that the retarded Green function $G_{\lambda \rho,\alpha
\beta}(k,t)$ of equation (A.2) satisfies the following equation:

\be
\Biggl [{\Large \bf{1}}_{\mu \nu, \lambda \rho} \frac{\partial}{\partial t} +
k^2({\Large \bf{1}} -{\Large \bf{P}})_{\mu \nu,\lambda \rho}\Biggr]
G_{\lambda \rho,\alpha \beta}(k,t) = {\Large \bf{1}}_{\mu \nu,\alpha \beta}
\delta(t).
\ee
Here we introduced the projectors

$$ {\Large \bf{1}}_{\mu
\nu,\lambda\rho} \equiv \frac{1}{2}\Biggl (\delta_{\mu \lambda} \delta_{\nu
\rho} - \delta_{\mu \rho} \delta_{\nu \lambda}\Biggr )~ \mbox{and} ~~~
{\Large \bf{P}}_{\mu \nu,\lambda \rho }\equiv \frac{1}{2}\Biggl (T_{\mu
\lambda} T_{\nu \rho} - T_{\mu \rho} T_{\nu \lambda}\Biggr) ,$$
where

$$
 T_{\mu \nu} \equiv \delta_{\mu \nu} - \frac{k_{\mu} k_{\nu}}{k^2}, $$
 possessing the following properties:

 \be
 {\Large \bf{1}}_{\mu \nu, \lambda
 \rho} ={- \Large \bf{1}}_{\nu\mu, \lambda \rho} = -{\Large \bf{1}}_{\mu
 \nu, \rho \lambda} = {\Large \bf{1}}_{\lambda \rho, \mu \nu}, ~~~ {\Large
 \bf{1}}_{\mu \nu, \lambda \rho} {\Large \bf{1}}_{\lambda \rho, \alpha
\beta} = {\Large \bf{1}}_{\mu \nu, \alpha \beta}
\ee
(the same properties
 hold true for ${\Large \bf{P}}_{\mu \nu, \lambda \rho}$),

$$
{\Large \bf{P}}_{\mu \nu, \lambda \rho}({\Large \bf{1}}
- {\Large \bf{P}})_{\lambda \rho, \alpha \beta} = 0.
$$
Using (A.4) and (A.5), one can
easily get $G_{\mu \nu, \alpha \beta}(k,t)$ from equation (A.3):
$$
G_{\mu \nu, \alpha
\beta}(k,t) = \theta(t) [({\Large \bf{1}} - {\Large \bf{P}}) e^{-k^2t} +
{\Large\bf{P}}]_{\mu \nu ,\alpha \beta}.
$$
Returning back to the coordinate
representation, one obtains, that the solution of equation (A.1) with the
initial condition $A_{\mu}(x,0) = A_{\mu}(x)$ has the form (7), where

$$
U_{\mu \nu, \beta}(y,t) = \frac{i}{8 \pi^4} \int dk~ e^{iky}
k_{\alpha} \Biggl [({\Large \bf{1}} - {\Large \bf{P}}) e^{-k^2t} + {\Large
\bf{P}} \Biggr ]_{\mu \nu, \beta \alpha},
$$
that after some calculations
yields (8).

\newpage
{\Large \bf{References}}

\noindent
1. {\it Parisi G. and Wu Y.}, Scienta Sinica 1981, vol.24,p.483;
for a review see

{\it Damgaard P.H. and H\"uffel H.}, Phys.Rep. 1987,
vol.152, pp.227--398.

\noindent
2. {\it Dosch H.G.}, Phys. Lett. B, 1987, vol. 190, p. 177;
{\it Simonov Yu.A.},

Nucl.Phys.B, 1988, vol.307,p.512;
{\it Dosch H.G. and Simonov Yu.A.},

Phys.Lett.B, 1988, vol.205,p.339.,Z.Phys.C, 1989, vol.45,p.147;

{\it Simonov Yu.A.}, Phys.Lett.B, 1989, vol.226,p.151.

\noindent
3. {\it Simonov Yu.A.}, Yad.Fiz., 1989,vol.50, p.213.

\noindent
4. {\it Simonov Yu.A.}, Nucl.Phys.B, 1989,vol.324,p.67.

\noindent
5. {\it Simonov Yu.A.}, Yad.Fiz.,1991, vol.54, p.192.

\noindent
6. {\it Badalyan A.M. and Yurov V.P.}, Yad.Fiz.,
1990,vol.51,p.1368.

\noindent
7. {\it Simonov Yu.A.}, Phys.Lett.B,1989,vol.228, p.413.

\noindent
8. {\it Simonov Yu.A.,} Phys.Lett.B, 1990, vol.249, p.514.

\noindent
9. {\it Van Kampen N.G.}, Stochastic processes in physics and
chemistry

(North-Holland Physics Publishing, Amsterdam,1984).

\noindent
10.{\it Antonov D.V. and Simonov Yu.A.}, Int.J.of Mod.Phys.A.
(in press).

\noindent
11.{\it Fradkin E.S. and Gitman D.M.}, Phys.Rev.D,
1991,vol.44,p.3230;

{\it Gitman D.M. and Shvartsman Sh.M.}, Phys.Lett.B, 1993, vol.318, p.122.

\noindent
12.{\it Fradkin E.S. et al.},Quantum electrodynamics with unstable
vacuum

(Springer--Verlag, Berlin, 1991).

\noindent
13.{\it Nielsen H.B. and Olesen P.}, Nucl.Phys. B, 1973,vol.61,p.45.

\noindent
14.{\it Breit J.D. et al.}, Nucl.Phys.B, 1984,vol.233, p.61;

{\it Damgaard P.H. and Tsokos K.}, Nucl.Phys.B, 1984, vol.235,p.75.

\noindent
15.{\it Grimus W. and Nardulli G.},Nuovo Cim.A, 1986,vol.91,p.384.

\noindent
16.{\it Weinberg S.}, Phys.Rev. D, 1974,vol.9,p.3357;

{\it Bernard C.W.}, Phys.Rev. D,1974,vol.9,p.3312;

{\it Dolan L. and Jackiw R.}, Phys.Rev.D, 1974, vol.9,p.3320.

\noindent
17.{\it Simonov Yu.A.}, hep-ph 9311216, Yad.Fiz., 1995, vol.58,p.357,

JETP Lett., 1992, vol.55,p.605;

{\it Gubankova E.L. and Simonov Yu.A.}, Phys.Lett. B, 1995, vol.360, p.93.

\noindent
18.{\it Vasiliev A.A.}, Functional methods in quantum field theory
and in statistics
(Leningrad State University, Leningrad,1976).
\end{document}